# Stimulated Globular Scattering of Laser Radiation in Photonic Crystals: Temperature Dependences

V. S. Gorelik (1), A. D. Kudryavtseva (1), N. V. Tcherniega (1), A. I. Vodchits (2)

((1) P. N. Lebedev Physical Institute, RAS, Moscow, Russia, (2) B. I. Stepanov Institute of Physics, National Academy of Sciences of Belarus, Minsk, Belarus)

## INTRODUCTION

The investigations of nonlinear optical effects are very important because they give information about material structure and light interaction with different kinds of substances and also they open wide field of applications. There are many methods of light signals processing, industrial production control, environment monitoring and others based on nonlinear effects, in particular, nonlinear optical effects in crystals. Recently many new crystals have been created for efficient use in lasers and nonlinear optical devices. Among them there are barium nitrate ($Ba(NO_3)_2$), barium tungstate ($BaWO_4$), potassium gadolinium tungstate ($KGd(WO_4)_2$ or KGW), potassium yttrium tungstate ($KY(WO_4)_2$ or KYW), potassium ytterbium tungstate ($KYb(WO_4)_2$ or KYbW). These crystals are considered as rather promising materials for Raman lasers based on stimulated Raman scattering (SRS) of light and some of them are extensively applied as laser host materials [1-4].

Nowadays new kind of nonlinear materials - photonic crystals - are being intensively studied because of their unique properties, which can give rise to the new class of optoelectronic devices [5-7]. Nano-electronics and nano-optics become very important for new technologies development. However, studying the nonlinear optical properties of nanomaterials has been started very recently. Therefore, there is no detailed information concerning these so far.

Very important kind of 3-D photonic crystals is globular photonic crystal built of globules (balls) with a diameter, which may be comparable with visible light wavelength. In nature such crystals exist as mineral - opal, consisting of quartz nanospheres. Space among these spheres (or globules) is filled with different inorganic materials. Recently technology of synthetic opals producing is developed [8-10]. Such opals have 3-D periodical structure and are built of ordered close-packed quartz globules with a diameter of 200-600 nm, organizing 3-D face-centered cubic lattice.

As the refractive index (n) contrast (refraction index ratio of $SiO_2$ and air, $n_{SiO2}/n_{air}$) in opal is about 1.45, the complete photonic band gap in such a structure does not exist but the photonic pseudo gap takes place. Empty cavities among these globules have octahedral and tetrahedral form. It is possible to investigate both initial opals (opal matrices) and nanocomposites, in which cavities are filled with organic or inorganic materials, for instance, semiconductors, superconductors, ferromagnetic substances, dielectrics, displaying different types of nonlinearities.

In the case of the different kinds of molecular liquids infiltrating the opal crystals some types of stimulated scatterings of light can be observed: SRS, stimulated Brillouin scattering (SBS) and others. Some peculiarities of spontaneous Raman scattering in photonic crystals has been described [11]. SBS in such structures was also observed [12]. Recently we observed very interesting nonlinear effects arising in photonic crystals (synthetic opals) under excitation with giant pulses of ruby laser: stimulated scattering in synthetic opals and opal nanocomposites - so-called stimulated globular scattering (SGS) [13] and new nonlinear effect - photonic flame effect (PFE) [14, 15]. SGS effect consisted in the appearance of one or two Stokes components shifted from the exciting light frequency by $0.4 – 0.6$ cm$^{-1}$. In the case of PFE we observed long-time radiation (with a temporal duration of few seconds) in the blue-green wavelength range under excitation of the sample with a 20 ns ruby laser pulse. The luminescence could appear with some delay in time in other opal crystals spatially separated from the crystal illuminated by the laser light.

Many nonlinear effects are known to change their properties substantially at low temperatures. For instance, conversion efficiency of the pumping light at SRS is increased with temperature decrease; self-focusing characteristics and others strongly depend on the material temperature. PFE appears only at the temperature of liquid nitrogen. Therefore, to clarify these effects, we have studied the temperature changes of the SGS and compared them with the temperature changes of SRS in the crystals of calcite. Below we present the results of these studies.

**EXPERIMENTAL**

Ruby laser giant pulses ($\lambda = 694.3$ nm, $\tau = 20$ ns, $E_{max} = 0.3$ J) were used to excite the nonlinear optical effects in photonic crystals and in calcite. Exciting light has been focused into the material by lenses with different focal lengths (50, 90, and 150 mm). The sample distance from focusing system and exciting light energy were also changed. It allowed us to vary the power density at the entrance of the sample and field distribution inside the sample. Investigations have been fulfilled for the different energy and geometrical conditions and for the different sample temperatures.

Fabri-Perot interferometers were used for SGS spectral structure investigations: the range of dispersion was changed from 0.42 to 1.67 cm$^{-1}$. The used samples of opal crystals had a size of 3×5×5 mm and were cut parallel to the plane (111). The angle of the incidence of the laser beam relative to the plane (111) was varied from 0 to 60$^0$. Opal crystals consisting of the close-packed amorphous spheres with a diameter of 200 nm and nanocomposites (opal crystals with voids filled with acetone or ethanol) were investigated. The experimental results did not depend on the angle of incidence of the laser beam. It is should be noted that opal matrix saturated with acetone or ethanol became practically transparent because refractive indices of such nanocomposite components (opal matrix and liquid) were approximately the same. It allowed observing the light scattering in forward direction.

We investigated SRS in calcite single crystal cut parallel to its axis [16]. The direction of the exciting beam coincided with the crystal axis direction. To prevent oscillation at the crystal (?) its entrance and exit surfaces were cut at an angle of 10$^0$. Spectral characteristics of the SRS in calcite were studied using a diffraction grating spectrograph (STE-1), energy parameters were measured with a sensitive photo-receiver (ELU-FT).

The optical scheme of the experimental setup for SGS study is shown in Fig. 1.

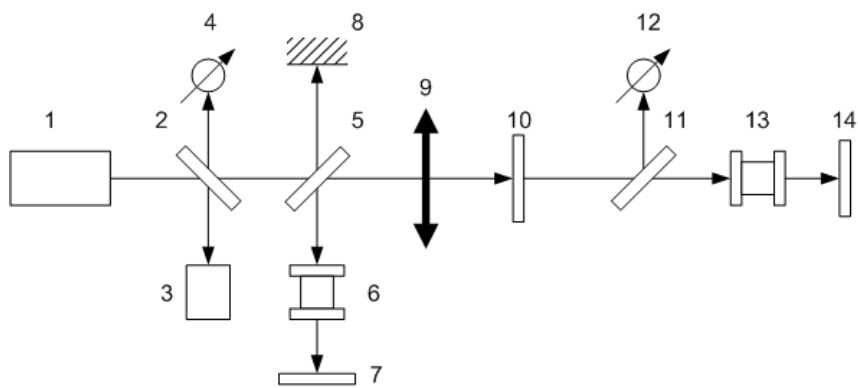

Fig. 1. Optical scheme of SGS studies: 1 - a ruby laser; 2, 5, 11 – glass plates; 3 – the system for laser parameters control; 4, 12 – the system for measuring the scattered light energy in backward and forward directions; 6, 13 – Fabri-Perot interferometers for SGS spectrum study; 7, 14 – the systems of spectra registration; 8 – a mirror; 9 – a lens; 10 – the studied crystal.

Ruby laser (1) light was focused into the crystal (10) with the lens (9). The system (3) was used for pumping light parameters control; the systems (4 and 12) measured stimulated scattering energy in backward and forward direction. Mirror (8) has been used to register initial light spectrum simultaneously with the scattered light spectrum. In some experiments it was put away. Fig. 2 shows an interferogram of the ruby laser oscillations. We can observe a pattern of rings, width of which characterizes a spectral width of the initial light. For our laser it was equal to 0.015 cm$^{-1}$.

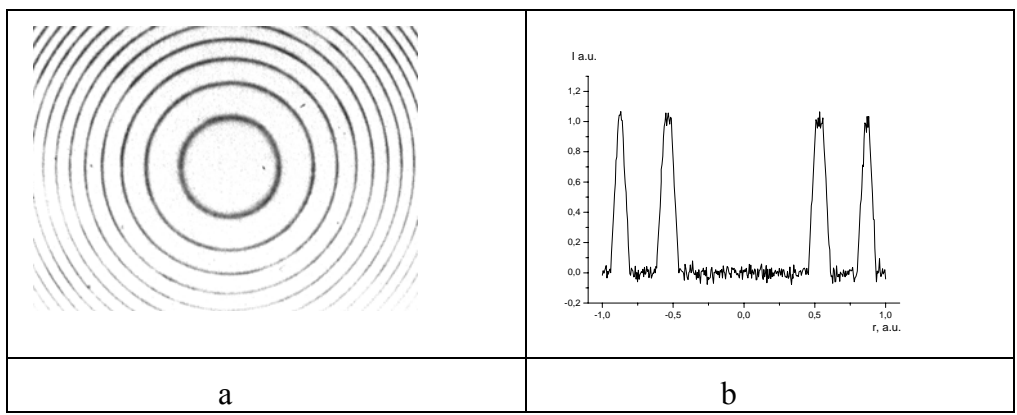

Fig. 2. Interferogram of the exciting laser light: a) - picture of the interferogram; b) - intensity distribution in the spectrum.

## RESULTS AND DISCUSSION

Under ruby laser giant pulse excitation we observed SGS both in opal matrix and in nanocomposites (opal matrix with pores filled with liquids). At room temperature, SGS in pure crystal was observed in backward direction (opposite to the initial pumping beam) for pumping light power density more than 0.12 GW/cm$^2$ and its frequency shift was 0.44 cm$^{-1}$. Energy conversion of the laser light into the SGS light was about 10 - 40 %. The divergence of the scattered light beam was of the order 10$^{-3}$ rad. The line width of the scattered light was close to the laser light line width.

We have also studied SGS in nanocomposites – opal crystals with pores between quartz globules filled with liquids: ethanol or acetone. We registered SGS in such structures both in forward and backward direction. In backward direction, for exciting light intensity more than 0.12 GW/cm$^2$ we observed the first Stokes component with a frequency shift of 0.4 cm$^{-1}$ in opals both with ethanol and acetone. Increasing the pumping power density resulted in the appearance of the second Stokes component with a frequency shift of 0.65 cm$^{-1}$ for acetone and 0.63 cm$^{-1}$ for ethanol. Spectra of backward SGS in acetone are shown in Fig. 3a and 3b.

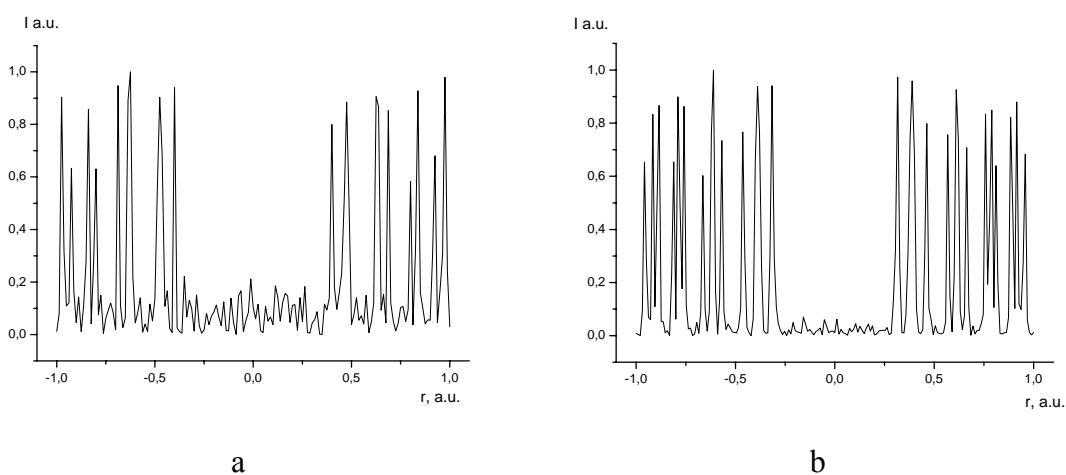

Fig. 3. Backward SGS spectrum for opal filled with acetone at room temperature. Dispersion range is 1.67 cm$^{-1}$.
a - pumping power density is 0.12 GW/cm$^2$, b - 0.21 GW/cm$^2$.

Larger ring corresponds to the laser light, smaller rings – to the SGS. Backward SGS spectrum in opal, filled with ethanol, is presented in Fig.5. In this spectrum, two Stokes SGS components are observed. Ring, corresponding to the laser light is not observed, because mirror 8 was put away for this case.

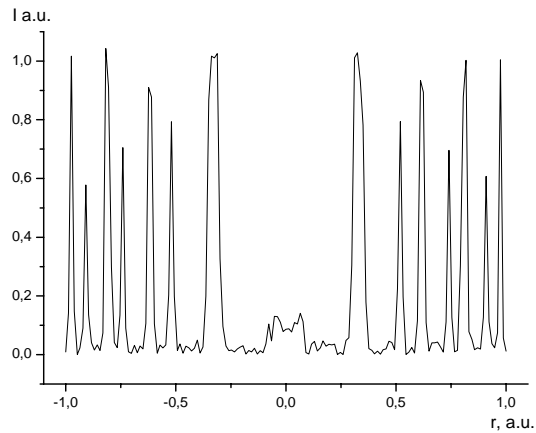

Fig. 4. Backward SGS spectrum for opal filled with ethanol at the room temperature for pumping power density 0.21 GW/cm$^2$. Range of dispersion is 1.67 cm$^{-1}$.

In forward direction, at the room temperature we observed only one Stokes component of SGS with a frequency shift of 0.4 cm$^{-1}$ in opal nanocomposites with ethanol and acetone. Forward SGS spectrum in opal filled with acetone is presented in Fig.5.

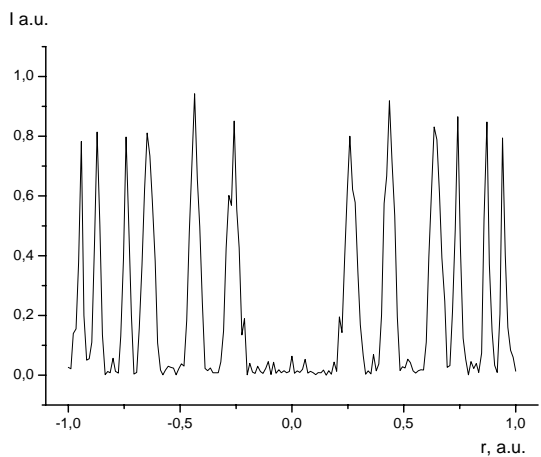

Fig. 5. SGS spectrum in the forward direction for opal crystal filled with acetone at room temperature. Range of dispersion is 1.67 cm$^{-1}$.

Ring with larger diameter corresponds to the laser light, with smaller diameter – to SGS. Spectrum of nanocomposite with ethanol is similar.

Decreasing the temperature leads to the SGS threshold lowering and to the arising of higher Stokes components. For liquid nitrogen temperature (77 K) SGS threshold is 0.04 GW/cm$^2$ (three times as small as one at the room temperature) and it is possible to observe 2 or 3 Stokes components in the forward direction, while at room temperature we observed only one Stokes component for the same experimental conditions. SGS spectrum in the forward direction in crystal filled with ethanol at liquid nitrogen temperature is presented in Fig. 6.

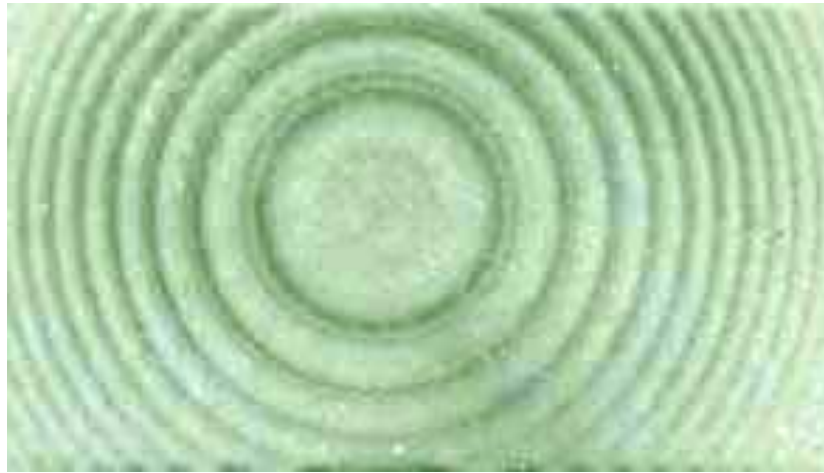

Fig. 6. SGS spectrum in the forward direction for opal crystal filled with ethanol at liquid nitrogen temperature. Range of dispersion is 1.67 cm$^{-1}$.

We compared our experimental results with eigenvalues of quartz globules vibrations theoretically. The equation of motion of such a body is

$$\rho \partial^2 \overline{D} / \partial t^2 = (\lambda + \mu)\overline{\nabla}(\overline{\nabla} \cdot \overline{D}) + \mu \overline{\nabla}^2 \overline{D}, \qquad (1)$$

where D is a displacement and the parameters $\lambda$ and $\mu$ are Lamé's constants. It is possible to solve this equation by introducing scalar and vector potentials. The scalar potential solution of the Helmholtz wave equation is

$$\phi_{bl} \propto Z_l(hr) P_l^m(\cos\theta) {{\cos m\varphi} \atop {\sin m\varphi}} \exp(-i\omega t), \qquad (2)$$

where $Z_l$ is a spherical Bessel function and h = $\omega$/c. The displacement derived from $\phi_{bl}$ becomes $\overline{D}_s = \overline{\nabla}\phi_s$. The vector potential is set as $\overline{A}$ = (r$\psi_v$, 0, 0), where

$$\psi_M \propto Z_l(kr) P_l^m(\cos\theta) {{\cos m\varphi} \atop {\sin m\varphi}} \exp(-i\omega t) \qquad (3)$$

with k = $\omega/c_l$.

Lamb obtained two types of modes under stress-free boundary condition at spherical surface. The torsional mode is forbidden by selection rules. The eigenvalue equation for spheroidal mode is derived from

$$2[\eta^2 + (l-1)(l+2)\{\eta j_{l+1}(\eta)/j_l(\eta) - (l+1)\}]\xi j_{l+1}(\xi)/j_l(\xi) - 0.5\eta^4 + (l-1)(2l+1)\eta^2 + \{\eta^2 - 2l(l-1)(l+2)\}\eta j_{l+1}(\eta)/j_l(\eta) = 0, \quad (4)$$

where the eigenvalues are

$$\xi = hR = \omega R/c_l = \pi\nu D/c_l \quad , \quad \eta = kR = \omega R/c_t = \pi\nu D/c_t, \quad (5)$$

and $j_l(\eta)$ is the first kind spherical Bessel function; $c_l$ and $c_t$ are longitudinal and transverse sound velocities. Equation (9) is solved by setting the parameter $c_l/c_t$. M.H. Kuok, H.S. Lim, S.C. Ng et al. [12] calculated frequencies $\nu$ for quartz spheres taking into account the values of longitudinal and transverse acoustic mode velocities: $c_l$ = 5279 m/s and $c_t$ = 3344 m/s. Calculated frequencies (in GHz) are following:

$$\nu_{10} = 2.617/D, \quad \nu_{12} = 2.796/D, \quad \nu_{20} = 4.017/D, \quad \nu_{30} = 6.343/D, \quad (6)$$

where D is sphere diameter (in $10^{-6}$ m). For our case (D = 200 nm) we have calculated the values of quartz globules frequencies: $\nu_{10}$ = 0.44 cm$^{-1}$, $\nu_{20}$ = 0.68 cm$^{-1}$, $\nu_{30}$ = 1.07 cm$^{-1}$, which are close to our experimental results. Experimental and calculated values of the frequencies are presented in Table 1.

Table 1. SGS Stokes frequencies for the different scattering geometries.

| Scattering geometry | $\nu$, cm$^{-1}$, experimental values | $\nu$, cm$^{-1}$, calculated values |
|---|---|---|
| Room temperature | | |
| Backward SGS in opal matrix | 0.44 | 0.44 |
| Backward SGS in nanocomposite (opal with acetone) | 0.40<br>0.65 | 0.44<br>0.68 |
| Backward SGS in nanocomposite (opal with ethanol) | 0.39<br>0.63 | 0.44<br>0.68 |
| Forward SGS in nanocomposite (opal with acetone) | 0.40 | 0.44 |
| Forward SGS in nanocomposite (opal with ethanol) | 0.37 | 0.44 |
| Liquid nitrogen temperature | | |
| Forward SGS in nanocomposite (opal with ethanol) | 0.40<br>0.77<br>1.13 | 0.44<br>0.68<br>1.07 |

We can see that good agreement exists for pure opal matrix. Liquids, filling pores between quartz globules, may influence their motion. Nevertheless, the difference between the experimental and calculated values of nanocomposites frequency shifts is not large even in this case.

It is interesting to compare the temperature dependence of the characteristics of SGS and SRS. We could register SRS near the threshold in calcite with a sensitive photo-receiver when the radiation intensity was 0.1 GW/cm$^2$, but its sharp growing began for the intensities equal to or higher 0.4 GW/cm$^2$. Comparing the thresholds for different kinds of scatterings is important to evaluate the role of different nonlinear effects and their competition in the processes, occurring in the material under laser light action. We investigated SRS in calcite for a wide temperature range: from +150 C to -196 C [16]. We found out that with the temperature decrease the threshold of SRS became much lower, energy was increased and redistributed into higher components. Conversion efficiency was also increased up to a saturation level which took place at the temperatures below than approximately -100 C.

Table 2 presents the temperature dependence of the 1$^{st}$ and 2$^{nd}$ Stokes components energy at SRS in calcite and total efficiency of SRS conversion of pumping light (it is a ratio of the energy of all observed SRS components in forward and backward directions to the exciting light energy).

Table 2. Energy of the SRS Stokes components (E) and conversion efficiency (η) of SRS in calcite for different temperatures.

| Temperature (C) | E·10$^3$ (J) | | η (%) |
| --- | --- | --- | --- |
| | 1 st | 2 st | |
| +150 | 57 | 0.003 | 23 |
| +20 | 68 | 0.008 | 33 |
| -20 | 126 | 0.017 | 56 |
| -110 | 197 | 0.016 | 76 |
| -196 | 177 | 0.590 | 76 |

One can see from this table that the first Stokes component energy at -196 C is about 3 times larger than at +150 C, and energy of the second Stokes component at -196 C is almost 200 times larger than at +150 C.

I.L. Fabelinski et al. have studied SBS in quartz at low temperatures and found that hypersound absorption is decreased with temperature lowering from 300 K to 80 K, which could lead to the SBS intensity increase. With temperature lowering to 4.2 K light canal became more narrow and as a result hypersound and SBS intensity were increased [17].

## CONCLUSIONS

These studies of temperature dependences of stimulated scatterings of laser radiation in different crystals (usual and photonic ones) show that some similar properties are observed for different types of crystals and for different kinds of stimulated scattering: increasing the overall stimulated scattering energy and its redistribution from the first (the most intensive) component to the higher order components. The efficiency of conversion of the exciting light into the stimulated scattered light is also increased reaching saturation at the definite temperature. Our previous investigations of the SRS in calcite [16] indicate that the first Stokes component intensity growth with the temperature lowering may be explained by the increase of the first Stokes component total intensity of the usual Raman scattering and its line width decrease with temperature lowering. Appearance of small-scale self-focusing at the liquid nitrogen temperature may also play definite role for conversion efficiency increase [18].

## ACKNOWLEDGEMENTS

The authors are grateful to the Russian Foundation for Basic Research for financial support (Grant N 06-02-81024-Bel_a).